\def\b{\begin{equation}}
\def\e{\end{equation}}
\def\balll{\begin{array}{lll}}
\def\ea{\end{array}}
\def\bea{\begin{eqnarray}}
\def\eea{\end{eqnarray}}

\documentclass[twocolumn,showpacs,preprintnumbers,amsmath,amssymb,eqsecnum,aps]{revtex4}
\usepackage{epsfig}
\begin{document}
\title{Semiclassical approach to black hole absorption of electromagnetic 
radiation emitted by a rotating charge}

\author{Jorge Casti\~neiras}
\email{jcastin@ufpa.br}
\author{Lu\'\i s C. B. Crispino}
\email{crispino@ufpa.br}
\affiliation{Departamento de F\'\i sica, Universidade Federal do
Par\'a, 66075-110, Bel\'em, PA,  Brazil}

\author{George E. A. Matsas}
\email{matsas@ift.unesp.br}
\affiliation{Instituto de F\'\i sica Te\'orica, Universidade Estadual Paulista, 
Rua Pamplona 145, 01405-900, S\~ao Paulo, SP, Brazil}

\author{Rodrigo Murta}
\email{murta@ufpa.br}
\affiliation{Departamento de F\'\i sica, Universidade
Federal do Par\'a, 66075-110, Bel\'em, PA,  Brazil}

\date{\today}
\begin{abstract}
We consider an electric charge, minimally coupled to the 
Maxwell field, rotating around a Schwarzschild black hole. 
We investigate how much of the radiation emitted from the swirling 
charge is absorbed by the black hole and show that 
most of the photons escape to infinity. For this purpose 
we use the Gupta-Bleuler quantization of the electromagnetic 
field in the modified Feynman gauge developed in the context of 
quantum field theory in Schwarzschild spacetime. 
We obtain that the two photon polarizations contribute quite 
differently to the emitted power. In addition, we discuss 
the accurateness of the results obtained in a full general 
relativistic approach in comparison with the ones obtained 
when the electric charge is assumed to be orbiting a massive object
due to a Newtonian force.  
\end{abstract}
\pacs{04.62.+v, 04.70.Dy, 41.60.-m}

\maketitle

\section{Introduction}
\label{sec:Introduction}

Much effort has been devoted to confirm 
the presence of black holes in X-ray binary systems \cite{evid1} 
as well as in galactic centers~\cite{evid2}. 
The analysis of the radiation emitted from accretion disks 
swirling around black hole candidates
may play a crucial role in the experimental confirmation 
of the existence of event horizons (see, e.g., Ref.~\cite{accdisks}).
It is interesting, thus, to compute the amount of the 
emitted radiation which is able to reach asymptotic observers rather 
than be absorbed by the hole. In a recent work, Higuchi and 
two of the authors analyzed the radiation 
emitted by a scalar source rotating around a 
Schwarzschild black hole~\cite{CHM2CQG}.
In the present work we study 
the more realistic case where the {\em scalar source} is
replaced by an {\em electric charge}. 
In order to capture the full influence
of the spacetime curvature on the emitted radiation, we work in the
context of quantum field theory in Schwarzschild spacetime 
(for a comprehensive account on quantum field theory in curved 
spacetimes 
see, e.g., Ref.~\cite{BD}). Because of the difficulty to express the
solution of some differential equations which we deal with 
in terms of known special functions
(see e.g. Ref.~\cite{Candetal} for a discussion on this issue) 
our computations are performed 
(i) numerically but without further approximations and 
(ii) analytically but restricted to the low-frequency regime 
(in which case the radial part of the normal modes
can be written in terms of Legendre functions~\cite{GR}).

We organize the paper as follows. In Section
\ref{sec:Quantization} we review the Gupta-Bleuler quantization 
of the electromagnetic field in a modified 
Feynman gauge in the spacetime of a static chargeless
black hole \cite{CHM3PRD}. We compute the radiated power from an electric charge 
swirling around a Schwarzschild black hole in Section \ref{sec:Rotating}.
In Section \ref{sec:Comparison} we compare this result with the one 
obtained considering the charge as orbiting a Newtonian object in flat spacetime.
Finally we use the previous results to compute in Section \ref{sec:Absorption}
what is the amount of the emitted radiation which is able to reach 
asymptotic observers. Our final remarks are made in Section \ref{sec:Final}.
We assume natural units $\hbar = c = G = 1$ and metric
signature $(+ - - -)$.

\section{Quantization of the electromagnetic field in 
Schwarzschild spacetime}
\label{sec:Quantization}

In this section we review the quantization of the massless
vector field in Schwarzschild spacetime following closely Ref. \cite{CHM3PRD}.
We write the line element of a static chargeless black hole as
\begin{equation}
 ds^{2} = f\left(  r\right)  dt^{2}-f\left(  r\right)
^{-1}dr^{2}-r^{2}d\theta^{2}-r^{2}\sin^{2}\theta d\phi^{2} \,,
\label{Schwliel}%
\end{equation}
where $f\left(  r\right)  =1-2M/r$. 

We then consider a massless vector field in this geometry
with classical action given by
\begin{equation}
S=\int d^{4}x
\mathcal{L} \, ,
\end{equation}
where the Lagrangian density in the modified Feynman gauge is
\begin{equation}
\mathcal{L}
=\sqrt{-g}\left[  -\dfrac{1}{4}F_{\mu\nu}F^{\mu\nu}-\dfrac{1}{2}G^{2}\right]
\label{Lagdens}%
\end{equation}
with $g=r^2 \sin \theta$, 
$G\equiv\nabla^{\mu}A_{\mu}+K^{\mu}A_{\mu}$ and 
\begin{equation}
K^{\mu}=\left(  0,{df}/{dr},0,0\right)  \text{.}%
\nonumber
\label{K}
\end{equation}
The corresponding Euler-Lagrange equations are, thus,
\begin{equation}
\nabla_{\nu}F^{\mu\nu}+\nabla^{\mu}G-K^{\mu}G=0\text{,}
\label{eqmo}%
\end{equation}
which can be cast in the form
\begin{eqnarray}
& & \dfrac{1}{f}\partial_{t}^{2}A_{t}-\dfrac{f}{r^{2}}\partial_{r}\left(
r^{2}\partial_{r}A_{t}\right)  +\dfrac{1}{r^{2}}\tilde{\nabla}^{2}%
A_{t}~=0 \,,
\label{eqt} \\
& & \dfrac{1}{f}\partial_{t}^{2}A_{r}-\dfrac{1}{f}\partial_{r}\left[
\dfrac{f^{2}}{r^{2}}\partial_{r}\left(  r^{2}A_{r}\right)  \right]  +\dfrac
{1}{r^{2}}\tilde{\nabla}^{2}A_{r}
\nonumber \\
& & +\dfrac{1}{f}\partial_{r}\left(  \dfrac
{f}{r^{2}}\right)  \tilde{\nabla}^{a}A_{a}=0\,,
\label{eqr} \\
& & \dfrac{1}{f}\partial_{t}^{2}A_{a}-\partial_{r}\left(  f~\partial_{r}%
A_{a}\right)  +\dfrac{1}{r^{2}}\left[  \tilde{\nabla}^{b}\left(  \tilde
{\nabla}_{b}A_{a}-\tilde{\nabla}_{a}A_{b}\right) \right.
\nonumber \\
& &  \left. +\partial_{a}\tilde{\nabla}^{b}A_{b}\right] 
+r^{2}\partial_{r}\left(  \dfrac{f}{r^{2}}\right)
\partial_{a}A_{r}=0\,.
\label{eqang}
\end{eqnarray}
Here $a$ and $b$ denote angular variables on the unit $2$-sphere $S^2$
with metric $\tilde{\eta}_{ab}$ and inverse metric $\tilde{\eta}^{ab}$ 
[with signature $(- -)]$, $\tilde{\nabla}_a$
is the associated covariant derivative on $S^2$,  
$\tilde{\nabla}^a \equiv \tilde{\eta}^{ab}\tilde{\nabla}_b$
and $\tilde{\nabla}^2 \equiv \tilde{\eta}_{ab} \tilde\nabla^a\tilde\nabla^b$.

We write the complete set of positive-frequency solutions of Eq. (\ref{eqmo})
with respect to the Killing field $\partial_t$ in the form
\begin{equation}
A^{(\varepsilon n \omega l m)}_{\mu} = 
\zeta^{\varepsilon n \omega l m}_{\mu}(r, \theta, \phi)
e^{-i\omega t}, \,\,\,\,\, \omega>0.
\label{KillSol}
\end{equation}
The index $\varepsilon$ stands for the four different polarizations.
The pure gauge modes, $\varepsilon = {\rm G}$, are the ones which satisfy
the gauge condition $G=0$ and can
be written as $A^{({\rm G} n \omega l m)}_{\mu} = \nabla_{\mu} \Lambda$,
where $\Lambda$ is a scalar field.
The physical modes, $\varepsilon = {\rm I}, {\rm II}$, satisfy the gauge condition
and are not pure gauge.
The nonphysical modes, $\varepsilon = {\rm NP}$, do not satisfy the gauge condition.
The modes incoming from the past null infinity ${\cal J}^-$ 
are denoted by $n = \leftarrow$
and the modes incoming from the past event horizon 
$H^{-}$ are denoted by $n = \rightarrow$.
The $l$ and $m$ are the angular momentum quantum numbers.

\begin{figure}[htb]
\vskip -2 truecm
\epsfig{file=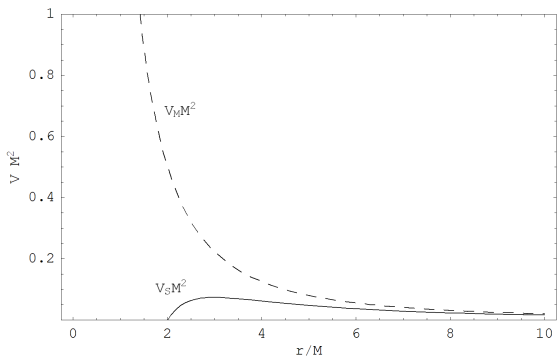,angle=0,width=0.999 \linewidth,clip=}
\vskip -2 truecm
\caption{
The potentials $V_S$ (solid line) and 
$V_M$ (dashed line), for $l=1$, are plotted as a function
of $r/M$ with $r$ being the radial coordinate in Schwarzschild and
Minkowski spacetimes, respectively. 
The Schwarzschild potential is restricted to $r>2M$
while for the Minkowski potential we have $r>0$.
Note that both $V_S$ and $V_M$ decrease asymptotically 
as $1/r^2$ but behave quite differently for $r\approx 2M$.
} 
\label{figure1}
\end{figure}

The physical modes can be written as
\begin{widetext}
\begin{equation}
A_{\mu}^{({\rm I}n\omega lm)}=\left( 0\,,\,
\dfrac{{\varphi}_{\omega l}^{{\rm I}n}\left(  r\right)}{r}  ~Y_{lm},\dfrac
{f}{l\left(  l+1\right)  }\dfrac{d}{dr}\left[  r{\varphi}_{\omega l}^{{\rm I}n}\left(
r\right)  \right]   
\partial_{\theta}Y_{lm},
\dfrac{f}{l\left(  l+1\right)  }\dfrac{d}{dr}\left[  r{\varphi}_{\omega l}%
^{{\rm I}n}\left(  r\right)  \right]  \partial_{\phi}Y_{lm} \right)e^{-i\omega t}%
\label{modo I}
\end{equation}
and
\begin{equation}
A_{\mu}^{({\rm II}n\omega lm)}=\left(  0,0,r{\varphi}_{\omega l}^{{\rm II}n}\left(  r\right)  Y_{\theta
}^{lm},r{\varphi}_{\omega l}^{{\rm II}n}\left(  r\right)  Y_{\phi}^{lm}\right)  e^{-i\omega
t} \label{modo II}%
\end{equation}
\end{widetext}
with $l\geqslant1$ (since the gauge condition $G=0$ is not 
satisfied for $l=0$).
The radial part of the physical modes satisfies 
the differential equation  
\begin{equation}
\left(  \omega^{2}-V_{S}\right)  \left[ r{\varphi}_{\omega l}^{\lambda n}\left(  r\right)\right]
+f\dfrac{d}{dr}\left(  f\dfrac{d}{dr}\left[ r{\varphi}_{\omega l}^{\lambda n}\left(
r\right)\right]  \right)  =0\text{,} \label{equacao em q}%
\end{equation}
where $\lambda = {\rm I}, {\rm II}$ and 
\begin{equation}
V_{S}=\left(  1-\dfrac{2M}{r}\right)  \dfrac{l\left(  l+1\right)  }{r^{2}%
}
\label{potencial em S}%
\end{equation}
is the Schwarzschild scattering potential (see solid line in Fig.~\ref{figure1}).
$Y_{lm}$ and $Y_a^{lm}$ are scalar and vector
spherical harmonics \cite{AHCQG}, respectively.
The remaining modes can be written as
\begin{equation}
A_{\mu}^{({\rm NP} n\omega lm)}=\left(  {\varphi}_{\omega l}^{{\rm NP} n}\left(  r\right)  Y_{lm}%
,0,0,0\right)  e^{-i\omega t} 
\label{modo NP }%
\end{equation}
and
\begin{equation}
A_{\mu}^{({\rm G} n\omega lm)}=\nabla_{\mu}\Lambda^{\omega nlm} \label{modo G}%
\end{equation}
with $l\geqslant0$, where
$$
\Lambda^{\omega nlm}=\dfrac{i}{\omega}{\varphi}_{\omega l}%
^{{\rm NP} n}\left(  r\right)  Y_{lm}e^{-i\omega t}
$$ 
and 
${\varphi}_{\omega l}^{{\rm NP} n}$ satisfies
\begin{equation}
\left(  \omega^{2}-V_{S}\right)  {\varphi}_{\omega l}^{{\rm NP} n}\left(  r\right)
+\dfrac{f^{2}}{r^{2}}\dfrac{d}{dr}\left(  r^{2}\dfrac{d}{dr}{\varphi}_{\omega l}%
^{{\rm NP} n}\left(  r\right)  \right)  =0\text{.} \label{equacao RNP }%
\end{equation}

The conjugate momenta associated with the field modes
are defined by
\begin{eqnarray}
\Pi^{(i)\mu\nu} & \equiv & \left.\dfrac{1}{\sqrt{-g}}\dfrac{\partial
\mathcal{L}}{\partial\left[  \nabla_{\mu}A_{\nu}\right]  }
\right| _ {A_\mu = A^{(i)}_{ \mu}}
\nonumber
\\
& = & -\left.\left[  F^{\mu\nu}+g^{\mu\nu
}G\right]\right| _ {A_\mu = A^{(i)}_{ \mu}}  \text{,} 
\label{momento S}
\end{eqnarray}
where $(i)$ represents $(\varepsilon, n, \omega, l, m)$.
By writing the conserved current
\begin{equation}
W^{\mu}\left[  A^{(i)},A^{(j)}\right]  \equiv i\left[  \overline{A_{\nu}^{(i)}}
\Pi^{(j)\mu\nu}-\overline{\Pi^{(i)\mu\nu}}A_{\nu}^{(j)}\right] \,,  
\label{corrente W}
\end{equation}
where the overline denotes complex conjugation,
we normalize the field modes through the generalized Klein-Gordon
inner product~\cite{CHM} defined by
\begin{equation}
\left(  A^{(i)},A^{(j)}\right)  \equiv\int_{\Sigma}
d\Sigma_{\mu}^{\left(  3\right)  }W^{\mu}\left[  A^{(i)},A^{(j)}\right]  \text{.}
\label{produto interno de KG generalizado}
\end{equation}
Here $d\Sigma_{\mu}^{\left(  3\right)  }\equiv d\Sigma^{\left(  3\right)
}~n_{\mu}$, where $d\Sigma^{\left(  3\right)  }$ 
is the invariant 3-volume element of the Cauchy surface $\Sigma$ and $n^{\mu}$ 
is the future pointing unit vector orthogonal to $\Sigma$.
The modes are then normalized such that
\begin{equation}
\left(  A^{(\varepsilon n \omega lm)},
        A^{(\varepsilon^{\prime} n^{\prime}\omega^{\prime
}l^{\prime}m^{\prime})}\right)  =M^{\varepsilon\varepsilon^{\prime}}%
\delta_{nn^{\prime}}\delta_{ll^{\prime}}\delta_{mm^{\prime}}\delta\left(
\omega-\omega^{\prime}\right)  \text{,}%
\end{equation}
where the matrix $M^{\varepsilon\varepsilon^{\prime}}$ is given by
\begin{equation}
M^{\varepsilon\varepsilon^{\prime}}=\left(
\begin{array}
[c]{cccc}%
1 & 0 & 0 & 0\\
0 & 1 & 0 & 0\\
0 & 0 & 0 & -1\\
0 & 0 & -1 & -1
\end{array}
\right)
\label{matriz}
\end{equation}
with $\varepsilon = ({\rm I} ,{\rm II}, {\rm G}, {\rm NP})$.

In order to quantize the electromagnetic field, we  
demand the equal time commutation relations
\begin{equation}
\left[  \hat{A}_{\mu}\left(  \mathbf{x},t\right)  ,\hat{A}_{\nu}\left(
\mathbf{x}^{\prime},t\right)  \right]  =\left[  \hat{\Pi}_{\mu t}\left(
\mathbf{x},t\right)  ,\hat{\Pi}_{\nu t}\left(  \mathbf{x}^{\prime},t\right)
\right]  =0\text{,} \label{comutador I}%
\end{equation}
\begin{equation}
\left[  \hat{A}_{\mu}\left(  \mathbf{x},t\right)  ,\hat{\Pi}^{t\nu}\left(
\mathbf{x}^{\prime},t\right)  \right]  =\dfrac{i\delta_{\mu}^{\nu}}{\sqrt{-g}%
}\delta^{3}\left(  \mathbf{x}-\mathbf{x}^{\prime}\right)  \text{.}
\label{comutador II}%
\end{equation}
The electromagnetic field operator can be expanded in terms of
the normal modes as
\begin{equation}
\hat{A}_{\mu}=\sum_{\varepsilon, n, l, m}\int_{0}^{\infty}d\omega\left[
\hat{a}_{( i) } A_{\mu}^{ (i) }
+\hat{a}_{{ (i) }  }^{\dagger}\overline{A_{\mu}^{ (i) }}\right] \,,
\label{expansao dos modos S}%
\end{equation}
where $\hat{a}_{( i) }$ and $\hat{a}_{{ (i) }  }^{\dagger}$
are the annihilation and creation operators, respectively, satisfying
\begin{equation}
\left[  \hat{a}_{(\varepsilon n \omega lm)},
\hat{a}^{\dagger}_{(\varepsilon^{\prime} n^{\prime}\omega^{\prime
}l^{\prime}m^{\prime})}\right]\!\!  
= \! (M^{-1})_{\varepsilon\varepsilon^{\prime}}%
\delta_{nn^{\prime}}\delta_{ll^{\prime}}\delta_{mm^{\prime}}\delta\left(
\omega-\omega^{\prime}\right).  
\end{equation}
The Fock space of the physical states $\left\vert {\rm PS}\right\rangle$
is obtained by imposing the Gupta-Bleuler condition~\cite{IZ}. In our case, this corresponds 
to impose
\begin{equation}
\hat{G}^{\left(  +\right)  }\left\vert {\rm PS}\right\rangle =0\text{,}
\label{GB}%
\end{equation}
where $\hat{G}^{\left(  +\right)}$ is the positive-frequency
part of the operator 
$\hat{G}=\nabla^{\mu}\hat{A}_{\mu}+K^{\mu}\hat{A}_{\mu}$. 
Condition (\ref{GB}) corresponds to
\begin{equation}
\hat{a}_{\left(  {\rm NP} \omega nlm\right)  }\left\vert {\rm PS}\right\rangle =0\text{.}
\label{operador no estado S}%
\end{equation}
The physical states are obtained by applying any number 
of creation operators $\hat{a}^{\dagger}_{({\rm I} n \omega lm)}$,
$\hat{a}^{\dagger}_{({\rm II} n \omega lm)}$ and
$\hat{a}^{\dagger}_{({\rm NP}  n \omega lm)}$
to the Boulware vacuum $\left\vert 0\right\rangle$ 
\cite{Boul} defined by
\begin{equation}
\hat{a}_{\left(  \varepsilon n\omega lm\right)  }
\left\vert 0\right\rangle =0\text{.}
\label{BoVa}%
\end{equation}
The creation operators associated with pure gauge
modes take physical states into nonphysical ones.
Moreover physical states of the form 
$\hat{a}^{\dagger}_{({\rm NP}  n \omega lm)}
\left\vert {\rm PS}\right\rangle$ have zero norm.
Therefore we can take as the representative elements 
of the Fock space those states obtained by applying
the creation operators associated with the two
physical modes to the Boulware vacuum.
For this reason we
will be concerned only with the two physical
modes, $\lambda = {\rm I}, {\rm II}$, in the rest of the paper.
(A more detailed discussion of the Gupta-Bleuler quantization
of the electromagnetic field in spherically symmetric and 
static spacetimes can be found in Ref.~\cite{CHM3PRD}.)

The solutions of Eq.~(\ref{equacao em q}) 
are functions whose properties are not well known. 
(See Ref.~\cite{Candetal} for some  properties.)
We can, however, obtain their analytic form (i) in the asymptotic 
regions for any frequency and (ii)
everywhere if we keep restricted to the low-frequency regime. 
In order to study the asymptotic behavior of the physical
modes we use the Wheeler coordinate
$x=\allowbreak r+2M\ln\left(  {r}/{2M}-1\right)  $
and rewrite Eq.~(\ref{equacao em q}) as
\begin{equation}
\left(  \omega^{2}-V_{S}\right)  [r{\varphi}_{\omega l}^{\lambda n}\left(  x\right)]
+\dfrac{d^{2}}{dx^{2}}\left[  r{\varphi}_{\omega l}^{\lambda n}\left(  x\right) \right]
 =0\text{.} \label{eq tipo Schrodinger}%
\end{equation}
Since the Schwarzschild potential~(\ref{potencial em S}) vanishes for
$r=2M$ and decreases as $1/r^2$ for $r \gg 2M$ (see Fig. \ref{figure1}),
the solutions of Eq. (\ref{eq tipo Schrodinger}) can be approximated 
in the asymptotic regions by
\begin{equation}
r{\varphi}_{\omega l}^{\lambda\rightarrow}\left(  r\right)  \approx\left\{
\begin{array}
[c]{c}
B_{\omega l}^{\lambda\rightarrow}\left(  e^{i\omega x}+R_{\omega l}
^{\lambda\rightarrow}e^{-i\omega x}\right) \\
B_{\omega l}^{\lambda\rightarrow}T_{\omega l}^{\lambda\rightarrow}
i^{l+1}\omega xh_{l}^{\left(  1\right)  }\left(  \omega x\right)
\end{array}
\right.
\begin{array}
[c]{c}
\left(  x\ll -1\right)  \text{,}\\
\left(  x\gg 1\right)  \text{,}
\end{array}
\label{solucao assintotica vai}
\end{equation}
and
\begin{equation}
r{\varphi}_{\omega l}^{\lambda\leftarrow}\left(  r\right)  \approx\left\{
\begin{array}
[c]{c}
B_{\omega l}^{\lambda\leftarrow}T_{\omega l}^{\lambda\leftarrow}e^{-i\omega
x}\\
B_{\omega l}^{\lambda\leftarrow}\left(  \left(  -i\right)  ^{l+1}\omega
xh_{l}^{\left(  1\right)  \ast}\left(  \omega x\right)  
\right.
\\
\left.
+ R_{\omega l}
^{\lambda\leftarrow}i^{l+1}\omega xh_{l}^{\left(  1\right)  }\left(  \omega
x\right)  \right)
\end{array}
\right.
\begin{array}
[c]{c}
\left(  x\ll -1\right)  \text{,}\\
\,  \\
\,  \\
\left(  x\gg 1\right)  \text{,}
\end{array}
\label{solucao assintotica vem}
\end{equation}
where 
$r{\varphi}_{\omega l}^{\lambda\rightarrow}\left(  r\right)$ 
and
$r{\varphi}_{\omega l}^{\lambda\leftarrow}\left(  r\right)$ 
are solutions incoming from $H^{-}$ and ${\cal J}^-$, 
respectively. Here
$h_{l}^{\left(  1\right)  }(x)$
is a spherical Bessel function of the third kind \cite{Abramo},
$B_{\omega l}^{\lambda n}$ are normalization constants, and
$
\left\vert R_{\omega l}^{\lambda n}\right\vert ^{2}~\text{and }\left\vert
T_{\omega l}^{\lambda n}\right\vert ^{2}
$
are the reflexion and transmission coefficients, respectively,
satisfying the usual probability conservation equation
$\left\vert R_{\omega l}^{\lambda n}\right\vert ^{2} + 
\left\vert T_{\omega l}^{\lambda n}\right\vert ^{2}= 1$.
Using the generalized Klein-Gordon inner product defined above
we obtain
\begin{equation}
\left\vert B_{\omega l}^{{\rm I}n}\right\vert  
 =  \sqrt{\dfrac{l\left(  l+1\right)}{4\pi}}
\,\omega^{-3/2}
\label{BI}
\end{equation}
and
\begin{equation}
\left\vert B_{\omega l}^{{\rm II}n}\right\vert  
 =  \dfrac{1}{\sqrt{4\pi}}\,\omega^{-1/2}\,\,\text{.}
\label{BII}
\end{equation}

Let us now find the analytic expressions of the physical modes
in the low-frequency approximation.
For this purpose we rewrite Eq. (\ref{equacao em q}) as
\begin{eqnarray}
& & \dfrac{d}{dz}\left[  \left(  1-z^{2}\right)  \dfrac{d{\varphi}_{\omega l}^{\lambda
n}\left(  z\right)  }{dz}\right]
\label{qz} 
\\
& & +\left[  l\left(  l+1\right)  -\dfrac
{2}{z+1}-\omega^{2}M^{2}\dfrac{\left(  z+1\right)  ^{3}}{z-1}\right]
{\varphi}_{\omega l}^{\lambda n}\left(  z\right)  =0\text{,}%
\nonumber
\end{eqnarray}
where $z\equiv {r}/{M}-1$. In the low frequency regime, 
we write the two independent solutions of Eq. (\ref{qz}) 
for $l\geqslant1$ as
\begin{equation}
{\varphi}_{\omega l}^{\lambda\rightarrow}\left(  z\right)  \approx C_{\omega
l}^{\lambda\rightarrow}\left[  Q_{l}\left(  z\right)  -\dfrac{\left(
z-1\right)  }{l\left(  l+1\right)  }\dfrac{dQ_{l}\left(  z\right)  }%
{dz}\right]  
\label{q Q}
\end{equation}
and
\begin{equation}
{\varphi}_{\omega l}^{\lambda\leftarrow}\left(  z\right)  \approx C_{\omega
l}^{\lambda\leftarrow}\left[  P_{l}\left(  z\right)  -\dfrac{\left(
z-1\right)  }{l\left(  l+1\right)  }\dfrac{dP_{l}\left(  z\right)  }%
{dz}\right]  \text{,} 
\label{q P}
\end{equation}
where $P_{l}\left(  z\right)  $ and $Q_{l}\left(  z\right)  $
are Legendre functions of the first and second kind~\cite{GR}, 
respectively,
and $C_{\omega l}^{\lambda n}$ are normalization constants.
We note that since 
$P_{l}\left(  z\right)  ~\approx~z^{l}$ 
and 
$Q_{l}\left(  z\right) ~\approx~z^{-l-1}$ for $z\gg 1$ 
and 
$P_{l}\left(  z\right)  \approx1$ 
and
$Q_{l}\left(  z\right)  \approx-\log\sqrt{z-1}$ for $z\approx1$, 
we obtain from Eqs. (\ref{q Q}) and 
(\ref{q P}) that ${\varphi}_{\omega l}^{\lambda\rightarrow}$ diverges in
$H^{-}$ and remains finite in ${\cal J}^-$, whereas ${\varphi}_{\omega l}
^{\lambda\leftarrow}$ diverges in ${\cal J}^-$ and remains finite in $H^{-}$. 
This is the reason why we have associated $Q_{l}\left(  z\right)  ~$ and
$P_{l}\left(  z\right)  $ with modes incoming from $H^{-}$ and 
${\cal J}^-$, respectively.

Now, by fitting asymptotically Eqs.~(\ref{q Q}) and~(\ref{q P}) 
with Eqs.~(\ref{solucao assintotica vai}) 
and~(\ref{solucao assintotica vem}), respectively,
we obtain that the 
normalization constants are (up to arbitrary phases) 
\begin{equation}
C_{\omega l}^{{\rm I}\rightarrow}=
2\,{\sqrt{\dfrac{l(l+1)}{\pi}}}\,\omega^{-1/2}\,\,,
\label{ctt I vai}
\end{equation}
\begin{equation}
C_{\omega l}^{{\rm II}\rightarrow}=
\dfrac{2}{{\sqrt{\pi}}}\,\omega^{1/2}\,\,,
\label{ctt II vai}
\end{equation}
\begin{equation}
C_{\omega l}^{{\rm I}\leftarrow}=\dfrac{1}{\sqrt{\pi l(l+1)}}\,
\dfrac{2^l \left( (l+1)!\right)^{2} M^{l}}
{\left(2l\right)!\left(  2l+1\right)!!}\,
\omega^{l-1/2} 
\label{ctt I vem}
\end{equation}
and
\begin{equation}
C_{\omega l}^{{\rm II}\leftarrow}=\dfrac{1}{\sqrt{\pi}}\,
\dfrac{2^l \left(l+1\right)\left(l!\right)^{2} M^{l}}
{l\left(2l\right)!\left(  2l+1\right)!!}\,
\omega^{l+1/2}\,\, \text{.} 
\label{ctt II vem}
\end{equation}

\section{Rotating charge in Schwarzschild spacetime}
\label{sec:Rotating}

Now let us consider an electric charge with $\theta=\pi/2$, 
$r=R_{S}$ and angular velocity $\Omega\equiv d\phi/dt= {\rm const} >0$ 
(as defined by asymptotic static observers), in uniform 
circular motion around a Schwarzschild black hole, described by 
the current density 
\begin{equation}
j_{S}^{\mu}\left(  x^{\nu}\right)  =\dfrac{q}{\sqrt{-g}u^{0}}\delta\left(
r-R_{S}\right)  \delta\left(  \theta-\pi/2\right)  \delta\left(
\varphi-\Omega t\right)  u^{\mu}\text{.} 
\label{jS}%
\end{equation}
Here $q$ is the coupling constant and 
\begin{equation}
u^{\mu}\left(  \Omega,R_{S}\right)  =
\dfrac{1}{\sqrt{f\left(
R_{S}\right)  -R_{S}^{2}\Omega^{2}}}\,
\left( 1 ,0,0,\Omega\right)
\label{uS}%
\end{equation}
is the charge's 4-velocity.
We note that $j_S^{\mu}$ is conserved, $\nabla_\mu j_S^\mu = 0$, and
thus 
$
\int_\Sigma d\Sigma_{\mu}^{\left(  3\right)  } j_S^{\mu}\left(  x^{\nu}\right)=q
$ 
for any Cauchy surface $\Sigma$.

Next let us minimally couple the charge to the field
through the action 
\begin{equation}
\hat{S}_{I}=
{\displaystyle\int}
d^{4}x\,\,\sqrt{-g}~j_S^{\mu} \hat{A}_{\mu} \text{.}
\label{acaoint}%
\end{equation}
Then the emission amplitude at the tree level of one 
photon with polarization $\varepsilon$ and quantum numbers 
$(n,\omega ,l,m)$ into the Boulware vacuum is given by
\begin{eqnarray}
\mathcal{A}^{\varepsilon n \omega lm}
& = & \left\langle \varepsilon n \omega lm \right\vert
i \hat{S}_{I}
\left\vert 0\right\rangle
\nonumber
\\
& = & i {\displaystyle\int}
d^{4}x\,\,\sqrt{-g}~j_S^{\mu}
\overline{A_{\mu}^{(\varepsilon n \omega lm)}}  \text{.} \label{emissao}%
\end{eqnarray} 
It can be shown that 
$\mathcal{A}^{\varepsilon n \omega lm}
\propto \delta \left( \omega - m\Omega\right)$.
This implies that only photons with frequency $\omega_0 = m\Omega$
are emitted once the charge has some fixed $\Omega = {\rm const}$.
One can also verify that the pure gauge and nonphysical modes 
have vanishing emission amplitudes. This is so for the pure gauge
modes because $\nabla_{\mu}j_S^{\mu}=0$ 
and for the nonphysical modes because they have zero norm.

The total emitted power is
\begin{equation}
W_{S}=\sum_{\lambda={\rm I},{\rm II}}\sum_{n=\leftarrow,\rightarrow}\sum_{l=1}^{\infty
}\sum_{m=1}^{l}\int_{0}^{+\infty}
d\omega~ \omega~\left\vert
\mathcal{A}^{\lambda n \omega lm}\right\vert ^{2}/T \text{,} 
\label{pot S}%
\end{equation}
where 
$T=2\pi\delta\left(  0\right)$  
is the total time as measured by the asymptotic static observers.
Using now Eqs.~(\ref{modo I})-(\ref{modo II}) and~(\ref{jS})-(\ref{uS})
we rewrite Eq.~(\ref{pot S}) as
\begin{equation}
W_{S}=\sum_{n=\leftarrow,\rightarrow}
\sum_{l=1}^{\infty}\sum_{m=1}^{l}
\left[ W_{S}^{{\rm I} n\omega_{0} lm}+W_{S}^{{\rm II} n\omega_{0} lm} \right] 
\label{pot Sch}
\end{equation}
with
\begin{eqnarray}
& & W_{S}^{{\rm I} n\omega_{0} lm}=
\dfrac{2\pi q^{2}m^3\Omega^3}
{\left[  l\left(  l+1\right)  \right]  ^{2}}
\left(  1-\dfrac{2M}{R_{S}}\right)  ^{2}
\nonumber
\\
& & \times\left[  \dfrac{d}{dR_{S}}\left[  R_{S}~{\varphi}_{\omega_{0} l}^{{\rm I}n}\left(
R_{S}\right)  \right]  \right]  ^{2}\left\vert Y_{lm}\left(  \pi/2,0\right)
\right\vert ^{2} 
\label{potencia IS}
\end{eqnarray}
and
\begin{eqnarray}
& & W_{S}^{{\rm II} n\omega_{0} lm}=
2\pi q^2 m\Omega^3 \left[  R_{S}~{\varphi}_{\omega_{0} l}
^{{\rm II}n}\left(  R_{S}\right)  \right]  ^{2}
\nonumber
\\
& & \times \left\vert Y^{lm}_\phi \left(  \pi/2,0\right)
\right\vert ^{2}\text{.} 
\label{potencia IIS}%
\end{eqnarray}
\begin{figure}
\vskip -2 truecm
\epsfig{file=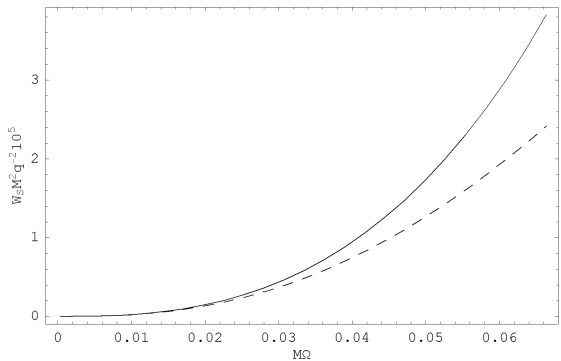,angle=0,width=0.999 \linewidth,clip=}
\vskip -2 truecm
\caption{The total power $W_S$ emitted by the electric
charge rotating around a Schwarzschild black hole
is plotted as a function of the angular velocity $\Omega$ 
as measured by asymptotic static observers.
The solid line represents our numerical result whereas
the dashed line represents our analytic result
for low frequencies. The $l$ summation in Eq. (\ref{pot Sch})
is performed up to $l=6$. $M \Omega$ ranges from $0$ up to
$0.068$ (associated with the innermost stable circular 
orbit at $R_S=6M$).} 
\label{figure2}
\end{figure}

Let us now relate the radial
coordinate $R_S$ of the rotating charge with its
angular velocity $\Omega$. According to General Relativity
for a stable circular orbit around a Schwarzschild black hole
we have \cite{Wald}
\begin{equation}
R_{S}=\left( {M}/{\Omega^{2}}\right)  ^{1/3}  \text{.} 
\label{RSW}%
\end{equation}
We use this relation to compute numerically the emitted power given by
Eqs.~(\ref{pot Sch})-(\ref{potencia IIS}) as a function of $\Omega$. 
The numerical method used here is analogous to the one
described in Ref.~\cite{CHM2CQG}.
The result is plotted as the solid line
in Fig.~\ref{figure2}. 
The main contribution to the emitted power comes from
modes with angular momentum $l=m=1$.
The larger the $l$,
the less is the contribution to the total radiated power.
For a fixed value of $l$, the dominant contribution comes from $m=l$.
Performing the summation up to $l=6$ in Eq.~(\ref{pot Sch}),
modes with $l=m=1$ give almost all the contribution
in the asymptotic region, while they contribute with
about 65\% at $R_S =6M$ (the innermost stable circular 
orbit according to General Relativity). In this case,  
the modes with $l=6$ contribute with less than 0.3\% of
the total radiated power for any position of the 
rotating charge.
\begin{figure}
\vskip -2 truecm
\epsfig{file=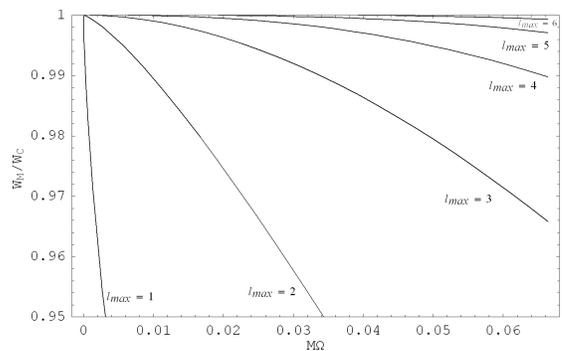,angle=0,width=0.999 \linewidth,clip=}
\vskip -2 truecm
\caption{The ratio $W_M/W_C$ is plotted as a function of
$\Omega$ considering an increasing number of contributions
of the angular momentum $l$. We perform the summation from
$l=1$ up to $l=l_{max}$ with $l_{max}$ varying from $1$
to $6$. We see that $W_M$ rapidly approaches the value 
of $W_C$ as we increase $l_{max}$. 
We also verify that it is important to consider 
large angular momentum contributions as the charge 
approaches the central object.}
\label{figure3}
\end{figure}

It is interesting to note that the magnitude of the
total radiated power in the electromagnetic case 
is approximately twice the numerical result found
previously for a scalar source coupled to a massless 
Klein-Gordon field \cite{CHM2CQG}. In principle,
this is not surprising because of the 
fact that photons have two physical polarizations.
Notwithstanding, it should be emphasized that 
the two polarizations 
contribute quite differently to the emitted power.
For our rotating charge,
the contribution from mode $\lambda= {\rm II}$ is negligible when
compared with the one from mode $\lambda= {\rm I}$  for every 
choice of $(n, \omega, l, m)$.
Considering angular momentum contributions up
to $l=6$, the ratio between the emitted power
associated with modes $\lambda= {\rm II}$ and 
$\lambda= {\rm I}$ is always less than 0.1\%.

Next, we use our low-frequency expressions for the 
physical modes, Eqs.~(\ref{q Q})-(\ref{ctt II vem}),
[and Eq.~(\ref{RSW})]
in Eqs.~(\ref{pot Sch})-(\ref{potencia IIS})
to obtain an analytic approximation for the emitted power. 
The result is plotted as the dashed line in Fig.~\ref{figure2}.
We see from it that the numerical and 
analytical  results differ sensibly as the charge approaches
the black hole but
coincide asymptotically, since far away 
from the hole only low frequency modes contribute to the 
emitted power.
\begin{figure}
\vskip -2 truecm
\epsfig{file=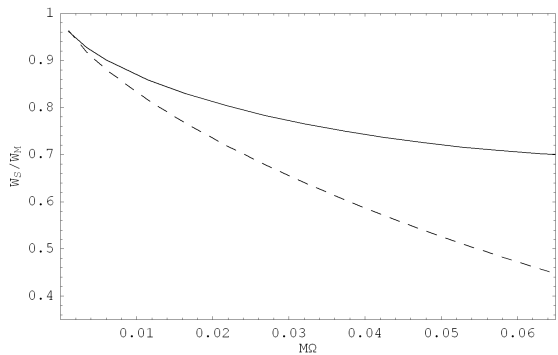,angle=0,width=0.999 \linewidth,clip=}
\vskip -2 truecm
\caption{The ratio $W_S/W_M$ is plotted as a function of
the angular velocity $\Omega$.
Again we consider contributions of the angular 
momentum up to $l=6$ in the summations in Eqs.~(\ref{pot Sch})
and~(\ref{potencia final M}). The maximum value of $M\Omega$ 
is 0.068. The solid line corresponds to our numerical result, 
while the dashed line corresponds to our analytical 
low-frequency approximation result.} 
\label{figure4}
\end{figure}

\section{Comparison with flat spacetime results}
\label{sec:Comparison}

In order to exhibit how better a full curved spacetime calculation 
can be in comparison with a flat spacetime one, let us show how the 
results found in the previous session differ from the ones obtained 
in Minkowski spacetime. In the latter case, the rotating charge is 
represented by the conserved current density
\begin{eqnarray}
j_M^{\mu}\left(  x^{\nu}\right)  & = & \dfrac{q}{R_{M}^{2}}\delta\left(
r-R_{M}\right)  \delta\left(  \theta-\pi/2\right)  \delta\left(  \phi-\Omega
t\right)  
\nonumber
\\
& \times & \left(  1,0,0,\Omega\right)  \text{.} \label{corrente}%
\end{eqnarray}
This is formally identical to Eq.~(\ref{jS}) but it is important 
to keep in mind that $R_S$ and $R_M$ are associated with Schwarzschild 
and Minkowski radial coordinates, respectively, which cannot be identified. 
The charge is regarded now as moving in a circular orbit 
due to a Newtonian gravitational force
around a central object with the same mass $M$ as the black hole. 
In order to relate
the radial coordinate $R_{M}$ with the angular 
velocity $\Omega$ (which is assumed to be measured by the same
asymptotic static observers as before), we use the Keplerian relation: 
$R_{M} = M^{1/3}\Omega^{-2/3}$.

The quantization of the electromagnetic field can be performed
analogously to the procedure exhibited in Section \ref{sec:Quantization}
by making $f = 1$. As a consequence, the scattering potential $V_S$
in Eq.~(\ref{potencial em S}) is replaced by 
$V_M = {l(l+1)}/{r^2}$. In Fig.~\ref{figure1} we plot the 
Minkowski scattering potential  for $l=1$ (see dashed line).

Assuming the same minimal coupling between the charge and the 
electromagnetic field as before,  we obtain that the emitted power 
at the tree level is given by
\begin{widetext}
\begin{eqnarray}
W_{M} &=& \frac{9 q^2 \Omega^{16/3} }{2} \sum_{l=1}^{\infty}\sum_{m=1}^{l}
\dfrac{m^{2} }{l(l+1)}
\left[  \dfrac{d}{d\Omega}\left[
\Omega^{-2/3}  ~j_{l}\left(  m\left(  \Omega
M\right)  ^{1/3}\right)  \right]  \right]  ^{2}\left\vert Y_{lm}\left(
\pi/2,0\right)  \right\vert ^{2}
\nonumber \\
&+& 
2 q^2 \Omega^{8/3} M^{2/3}
\sum_{l=1}^{\infty}\sum_{m=1}^{l}
 m^{2} \left[
j_{l}\left(  m\left(  \Omega M\right)  ^{1/3}\right)  \right]  ^{2}\left\vert
Y_{\phi}^{lm}\left(  \pi/2,0\right)  \right\vert ^{2}
\,. 
\label{potencia final M}%
\end{eqnarray}
\end{widetext}
The modes responsible for the dominant contributions
to the radiated power follow the same pattern as in 
the Schwarzschild case. In particular, the contribution 
from the physical modes $\lambda= {\rm II}$ is negligible 
when compared with the contribution from the 
physical modes $\lambda= {\rm I}$.

As a consistency check for the flat spacetime results, we compare
our quantum-oriented calculations with classical-oriented ones which
lead to the Larmor formula for the total emitted power. Applying it to
the case of a Keplerian circular orbit, we obtain
\begin{equation}
W_{C}=\dfrac{q^{2}M^{2/3}\Omega^{8/3}\gamma_{M}^{4}}
{6\pi} 
\label{pot green}%
\end{equation}
with
$
\gamma_{M}=(1- {M}^{2/3}\Omega^{2/3})^{-1/2}\text{.} 
$
In Fig.~\ref{figure3} we plot the ratios between $W_{M}$ and $W_{C}$,
where the summations in Eq.~(\ref{potencia final M}) are performed 
up to increasing values of $l=l_{\rm max}$. 
We see that for $l_{max}=6$ the difference between 
$W_{M}$ and $W_{C}$ is less than 0.1\%
for $R_M>6M$. This is in agreement with 
the fact that the contributions associated 
with higher values of $l$ are negligible to the emitted power. 

Next we compare our curved and flat spacetime results using 
our previous expressions for $W_S$ and $W_M$
as functions of the physical observables $M$ and $\Omega$ 
as measured by asymptotic static observers.
We plot the ratio between $W_S$ and $W_M$ in Fig.~\ref{figure4}
obtained from our numerical computations (solid line) and from
our low-frequency analytic approximation (dashed line).
In both cases the ratio tends to the unity as the charge rotates
far away from the attractive center, as a consequence 
of the fact that the Schwarzschild spacetime is asymptotically 
flat. As the rotating charge approaches the central object,
curved and flat spacetime results differ more significantly.
In the innermost relativistic stable circular 
orbit, the numerical computation gives that $W_S$ is 30\% 
smaller than $W_M$. We emphasize that this is not a simple 
consequence of the red-shift effect, since
the mode functions representing the quanta of the emitted 
radiation are quite different in curved and
flat spacetimes.

\section{Absorption of the electromagnetic radiation by 
the black hole}
\label{sec:Absorption}

\begin{figure}
\vskip -2 truecm
\epsfig{file=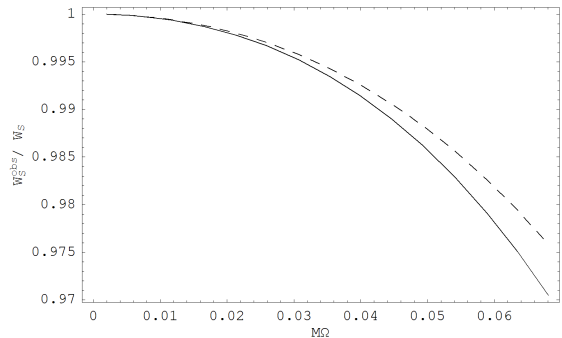,angle=0,width=0.999 \linewidth,clip=}
\vskip -2 truecm
\caption{The ratio between the asymptotically observed power 
$W_{S}^{obs}$ and the emitted power $W_S$ is plotted as a function
of the angular velocity $\Omega$ of the swirling electric
charge according to asymptotic static observers. 
$M\Omega$ varies from 0 to $0.068$. The summations in Eqs.~(\ref{pot Obs})
and~(\ref{pot Sch}) are performed up to $l=6$.} 
\label{figure5}
\end{figure}

Now, this is interesting to use our quantum field theory in  
Schwarzschild spacetime approach to compute what is the amount of emitted
radiation  which can be asymptotically observed. This is given by 
\begin{eqnarray}
W_{S}^{obs}=\sum_{\lambda={\rm I},{\rm II}}
\sum_{l=1}^{\infty}\sum_{m=1}^{l}
\left[ \vert T_{\omega_0 l}^{\lambda\rightarrow} \vert^2 
W_{S}^{\lambda \rightarrow \omega_{0} lm}\right.
\nonumber
\\
\left. + \vert R_{\omega_0 l}^{\lambda\leftarrow} \vert^2
W_{S}^{\lambda \leftarrow \omega_{0} lm} \right]\,.
\label{pot Obs}%
\end{eqnarray}
Our numerical result is shown as the solid line in 
Fig.~\ref{figure5}. 
In order to compute $W_{S}^{obs}$ in the low-frequency
approximation (see dashed line in Fig.~\ref{figure5}) 
we calculate the expressions for
the transmission and reflection coefficients.
For this purpose we use 
\begin{equation}
Q_{l}\approx\dfrac{2^{l}\left(  l!\right)  ^{2}}{\left(  2l+1\right)
!}z^{-l-1}\,\,\,\,\,\left(z\gg 1\right)
\label{Qsmallz}
\end{equation}
in Eq.~(\ref{q Q}) to write
\begin{equation}
r{\varphi}_{\omega l}^{\lambda\rightarrow}  \approx C_{\omega
l}^{\lambda\rightarrow}\,\dfrac{2^{l}\left(  l+1\right)  \left(  l!\right)
^{2}M^{l+1}}{l\left(  2l+1\right)  !}\,x^{-l}
\,\,\,\,\,\left(x\gg 1\right)\,\,\text{.}%
\label{rqsmallz}
\end{equation}
Comparing the above expression with Eq. (\ref{solucao assintotica vai}) and
using that $h_{l}^{\left(  1\right)  }\left(  x\right)  \approx\eta_{l}\left(
x\right)  \approx\dfrac{-\left(  2l\right)  !}{2^{l}l!}x^{-l-1}$ 
for $x\gg 1$, we obtain
\begin{equation}
\vert T_{\omega l}^{\lambda\rightarrow} \vert =
\dfrac{\vert C_{\omega l}^{\lambda\rightarrow}\vert}{\vert B_{\omega l}
^{\lambda\rightarrow}\vert}\,
\dfrac{2^{2l}\left(  l+1\right)\left(  l!\right)^{3}
M^{l+1}\omega^{l}}{l\left( 2l\right)!\left(
2l+1\right)  !}
\label{Tsmallz}
\end{equation}
with the normalization constants given by Eqs. (\ref{BI}),
(\ref{BII}), (\ref{ctt I vai}) and (\ref{ctt II vai}).
The reflexion coefficients are determined using that
$\left\vert R_{\omega l}^{\lambda\leftarrow}\right\vert
^{2}=\left\vert R_{\omega l}^{\lambda\rightarrow}\right\vert ^{2}=1-\left\vert
T_{\omega l}^{\lambda\rightarrow}\right\vert ^{2}$.
We see from Fig. \ref{figure5} that the black hole absorbs 
only a small amount of the emitted radiation. Even for the innermost 
stable circular orbit the black hole absorbs only $3\%$ of the 
total radiated power. These results are consistent with the fact 
that the absorption cross 
section of a Schwarzschild black hole is proportional to
$\omega^2$ for small frequency photons
(see, e.g., \cite{Page}).

\section{Final Remarks}
\label{sec:Final}

In this paper we have considered the radiation emitted by an
electric charge rotating around a chargeless static black 
hole in the context of quantum field theory in curved spacetimes. 
We have obtained that the two physical photon
polarizations give very different contributions 
to the total emitted power. Indeed, the contribution of one 
of the physical modes is negligible as compared with the other one.
As a consistency check of our procedure we have computed, using a
similar approach, the emitted power from a charge in Minkowski 
spacetime rotating around a massive object due to a Newtonian force
and showed that this is in agreement with Larmor's classical result.
Then we compared the radiation emitted (as measured 
by asymptotic static observers)
considering the attractive central object with mass $M$ 
as (i) a Schwarzschild black hole and (ii) a Newtonian 
massive object in flat spacetime. We have obtained that curved and flat 
spacetime results coincide when the charge orbits far away from the massive
object but differ considerably when the charge orbits close to it. 
The difference reaches 30\% for the
innermost stable circular orbit.
This result corroborates the importance of considering the curvature
of the spacetime in astrophysical phenomena occurring in the 
vicinity of black holes when they involve particles with wavelengths 
of the order of the event horizon radius.
Finally, we have computed the amount of the emitted radiation which
is absorbed by the black hole.
We have shown that most of the emitted radiation 
can be asymptotically observed. For the case of the innermost stable 
circular orbit at  $R_S=6M$, about 97\% of the emitted power
can be in principle detected at infinity.

\acknowledgments

The authors are grateful to Conselho Nacional de Desenvolvimento 
Cient\'\i fico e Tecnol\'ogico (CNPq) for partial financial support.
R. M. and G. M. would like to acknowledge also partial financial support
from Coordena\c{c}\~ao de Aperfei\c{c}oamento de Pessoal
de N\'\i vel Superior (CAPES) and Funda\c{c}\~ao de Amparo \`a
Pesquisa do Estado de S\~ao Paulo (FAPESP), respectively.

\end{document}